# An Active Defense Mechanism for TCP SYN flooding attacks

Saravanan K[1], Gowri Shankar A[2]

1ME (CSE) Student, MCET Pollachi, India

2Asst.Professor, MCET Pollachi, India

Email:saravanankumarasamy@gmail.com[1], ags@drmcet.ac.in[2]

## Abstract

*Distributed denial-of-service attacks on public servers have recently become a serious problem. To assure that network services will not be interrupted and more effective defense mechanisms to protect against malicious traffic, especially SYN floods. One problem in detecting SYN flood traffic is that server nodes or firewalls cannot distinguish the SYN packets of normal TCP connections from those of a SYN flood attack. Another problem is single-point defenses (e.g. firewalls) lack the scalability needed to handle an increase in the attack traffic. We have designed a new defense mechanism to detect the SYN flood attacks. First, we introduce a mechanism for detecting SYN flood traffic more accurately by taking into consideration the time variation of arrival traffic. We investigate the statistics regarding the arrival rates of both normal TCP SYN packets and SYN flood attack packets. We then describe a new detection mechanism based on these statistics. Through the trace driven approach defense nodes which receive the alert messages can identify legitimate traffic and block malicious traffic by delegating SYN/ACK packets.*

# An Active Defense Mechanism for TCP SYN flooding attacks

## 1. INTRODUCTION

On the Internet, a distributed denial-of-service (DDoS) attack is one in which a multitude of compromised systems attack a single target, thereby causing denial of service for users of the targeted system. The flood of incoming messages to the target system essentially forces it to shut down, thereby denying service to the system to legitimate users. An attempt to make a computer resource unavailable to its intended users. Although the means to, motives for and targets of a DoS attack may vary, it generally comprises the concerted, malevolent efforts of a person or persons to prevent an Internet site or service from functioning efficiently or at all, temporarily or indefinitely. It has been shown that more than 90% of the DoS attacks use TCP [1]. The TCP SYN flooding is the most commonly-used attack [2]. It consists of a stream of spoofed TCP SYN packets directed to a listening TCP port of the victim. Not only the Web servers but also any system connected to the Internet providing TCP-based network services, such as FTP servers or Mail servers, is susceptible to the TCP SYN flooding attacks. SYN attacks exploits TCP's three-way handshake mechanism and TCP's limitation in maintaining half-open connections.

## 2. CLASSIFICATION OF DOS ATTACKS

There are several general categories of DoS attacks. They are classified into three types: bandwidth attacks, logic attacks, and protocol attacks.

### 2.1 Bandwidth attacks
Bandwidth attacks are relatively straightforward attempts to consume resources, such as network bandwidth or equipment throughput. High-data-volume attacks can consume all available bandwidth between an ISP and site. The link fills up, and legitimate traffic slows down. Timeouts may occur, causing retransmission, generating even more traffic. An attacker can consume bandwidth by transmitting any traffic at all the network connection. [17]. A basic flood attack might use UDP or ICMP packets to simply consume all available bandwidth.

### 2.2 Logic Attacks
logic attacks exploit vulnerabilities in network software, such as a web server, or the underlying TCP/IP stack. Some vulnerability by crafting even a single malformed packet. They following are few examples logical attacks. Teardrop attacks sending IP fragments with overlapping, over-sized, payloads to the target machine. Peer-to-peer attacks have found a way to exploit a number of bugs in peer-to-peer servers to initiate DDoS attacks. Application level floods are Various DoS-causing exploits such as **buffer overflow** can cause server-running software to get confused and fill the disk space or consume all available memory or CPU time. A Nuke is an old denial-of-service attack against computer networks consisting of fragmented or otherwise invalid ICMP packets to the target, achieved by using a modified ping utility to repeatedly send this corrupt data, thus slowing down the affected computer until it comes to a complete stop

### 2.3 Protocol Attacks
The basic flood attack can be further refined to take advantage of the inherent design of common network protocols. These attacks do not directly exploit weaknesses in TCP/IP stacks or network applications but, instead, use the expected behavior of protocols such as TCP, UDP, and ICMP to the attacker's advantage. Examples of protocol attacks SYN flood is an asymmetric resource starvation attack in which the attacker floods the victim with TCP SYN packets and the victim allocates resources to accept perceived incoming connection. These are classified as Smurf Attack [3], SYN attack, UDP Attack, ICMP Attack, CGI request attack, Authentication server attack, Attack using DNS systems, Attack using spoofed address in ping.

# An Active Defense Mechanism for TCP SYN flooding attacks

*2.3.1Smurf Attacks*

The network floods it with excessive messages in order to impede normal traffic. It is accomplished by sending ping requests (ICMP echo requests) to a broadcast address on the target network or an intermediate network. The return address is spoofed to the Victim's address

*2.3.2UDP Attacks*

Using UDP for denial-of-service attacks is not as straightforward as with the TCP.The UDP flood attack can be initiated by sending a large number of UDP packets to random ports on a remote host. Thus, for a large number of UDP packets, the victimized system will be forced into sending many ICMP packets, eventually leading it to be unreachable by other clients. The attacker may also spoof the IP address of the UDP packets.

*2.3.3SYN Flood Attack*

A **SYN flood** is a form of denial-of-service attack in which an attacker sends a succession of SYN requests to a target's system. When a client attempts to start a TCP connection to a server, the client and server exchange a series of messages which normally working the client requests a connection by sending a SYN (synchronize) message to the server. The server acknowledges this request by sending SYN-ACK back to the client. The client responds with an ACK, and the connection is established. This is called the TCP three-way handshake, and is the foundation for every connection established using the TCP protocol. This is a well known type of attack and is generally not effective against modern networks. It works if a server allocates resources after receiving a SYN, but before it has received the ACK.

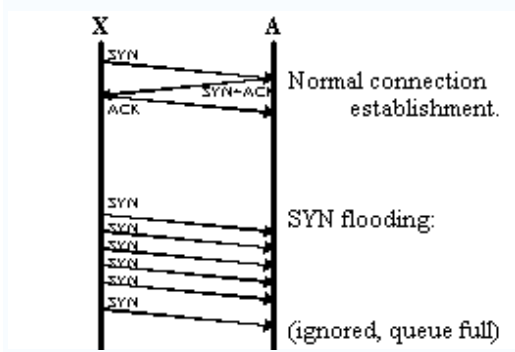

## 3. PREVENTION AND RESPONSE

### 3.1 SYN Cache

In the SYN cache [4] mechanism, the server node has a global hash table to keep half-open states of all applications, while in the original TCP these are stored in the backlog queue provided for each application. As a result, the node can have a larger number of half-open states and the impact of a SYN flood attack can be reduced.

### 3.2SYN Cookies

SYN cookies [5] modify the TCP protocol handling of the server by delaying allocation of resources until the client address has been verified. This seems to be the most powerful defense against SYN attacks. This technique used to guard against SYN flood attacks. The use of SYN Cookies [15] allows a server to avoid dropping connections when the SYN queue fills up. Instead, the server behaves as if the SYN queue had been enlarged. The server sends back the appropriate SYN+ACK response to the client but discards the SYN queue entry. If the server then receives a subsequent ACK response from the client, the server is able to reconstruct the SYN queue entry using information encoded in the TCP sequence number.

### 3.3Firewalls

Firewalls have simple rules such as to allow or deny protocols, ports or IP addresses. Some DoS attacks are too complex for today's firewalls, e.g. if there is an attack on port 80 (web service), firewalls cannot prevent that attack because they cannot distinguish good traffic from DoS attack traffic. Additionally, firewalls are too deep in the network hierarchy. The router may be affected even before the firewall gets the traffic. Nonetheless, firewalls can effectively prevent users from launching simple flooding type attacks from machines behind the firewall.

### 3.4Switches and Routers

Most switches have some rate-limiting and ACL capability. Some switches provide automatic and or system-wide rate limiting, traffic shaping, delayed binding to detect and remediate denial of service attacks through automatic rate filtering and WAN Link failover and balancing.

# An Active Defense Mechanism for TCP SYN flooding attacks

These schemes will work as long as the DoS attacks are something that can be prevented using them. For example SYN flood can be prevented using delayed binding or TCP splicing. Similarly content based DoS can be prevented using deep packet inspection. Attacks originating from dark addresses or going to dark addresses can be prevented using Bogon filtering. Automatic rate filtering can work as long as fixed the set rate-thresholds correctly and granularly. Wan-link failover will work as long as both links have DoS/DDoS prevention mechanism [11].

## 4. METHODOLOGY

In the exiting system solution for SYN attack is provided only based on SYN packets (i.e., the head of the connection) and FIN/RST packets (i.e., the tail of the connection) [11]. If the rate of SYN packets is much higher than that of FIN or RST packets, the router recognizes that attacking traffic is mixed in with the current traffic. Attacks can also be detected through the number of source addresses [15]. If the number of source addresses increases rapidly, the current traffic might include attack packets. These methods have several problems, however, one of which is that they cannot detect attacks until servers are seriously damaged or until most of the connections are closed. Another is that they may mistake high-rate normal traffic for attack traffic because they do not take into consideration the normal time-of-day variation of network traffic or they do so using a non-parametric approach without knowing how normal traffic varies. A non-parametric approach can detect attacks if there is any variance from normal traffic, but require a long time. Attack traffic should be identified more accurately and quickly by considering the variance of normal traffic. So, we propose a new defense mechanism for TCP SYN Flooding Attack.

### 4.1 Proposed Defense mechanism

TCP is a connection-oriented protocol. It uses various flags to indicate that a connection is being started or ended, or that the data carries a high priority. Many attacks are based on altering the TCP flags. Certain illegal combinations of TCP flags may be able to help packets avoid detection by firewalls or intrusion detection systems; other illegal combinations may be used to attack the systems. The functional specification for TCP is defined in RFC 793. This RFC and others define how systems should respond to legitimate packets, but they don't explain how systems should handle illegal combinations of flags. Consequently, different operating systems respond differently to illegal flag combinations. Attackers can exploit this to determine what operating system a device is using.

At least one of these six flags must be set in each TCP packet, each flag corresponds to a particular bit in the TCP header. The six flags [6] are

- SYN (Synchronization) - Initiate a TCP connection.
- ACK (Acknowledgment) - Indicates that the value in the acknowledgment number field is valid.
- FIN (Finish) - Gracefully end a TCP connection.
- RST (Reset) - Immediately end a TCP connection.
- PSH (Push) - Tells the receiver to pass on the data as soon as possible.
- URG (Urgent) - Indicates that the urgent pointer is valid; often caused by an interrupt.

*Case 1: Detecting by using TCP flags*

Normally connection between client and server are take place SYN, SYN ACK, and ACK are used during the three-way handshake which establishes a TCP connection between the client and the server. Except for the initial SYN packet, every packet in a connection must have the ACK bit set. FIN ACK and ACK are used during the graceful teardown of an existing connection. PSH FIN ACK may also be seen at the beginning of a graceful teardown. RST or RST ACK can be used to immediately terminate an existing connection. Packets during the "conversation" portion of the connection (after the three-way handshake but before the teardown or termination) contain just an ACK by default. Optionally, they may also contain PSH and/or URG. Packets with any other flag combination can be classified as abnormal. Here are some of the most commonly occurring are

- SYN FIN is probably the best known illegal combination. The SYN is used to start a connection, while FIN is used to end an existing connection. It is nonsensical to perform both actions at the same time. Many scanning tools use SYN FIN packets, because

# An Active Defense Mechanism for TCP SYN flooding attacks

many intrusion detection systems did not catch these in the past, although most do so now. You can safely assume that any SYN FIN packets you see are malicious.

- SYN FIN PSH, SYN FIN RST, SYN FIN RST PSH, and other variants on SYN FIN also exist. These packets may be used by attackers who are aware that intrusion detection systems may be looking for packets with just the SYN and FIN bits set, not additional bits set. Again, these are clearly malicious.

*Case 2: Detecting by using port*

There are several other characteristics of TCP traffic where abnormalities may be occurred to attackers Packets should never have a source or destination port set to 0. The acknowledgment number should never be set to 0 when the ACK flag is set. A SYN only packet, which should only occur when a new connection is being initiated, should not contain any data.

*Case 3: Detecting by ICMP feedback*

The hacker may send SYN packet to the server using some other IP address like 10.1.5.20. Then the server receives SYN packet and acknowledges to the corresponding IP address. But the client does not understand the SYN+ACK packet. And also, if the client is in off, the server will not receive the ACK packet and then waiting for the reply.

Our implementation takes place when the server replies SYN+ACK, the ICMP messages also added with that and it is sent to the client. Our approach is used to get the information like whether the client receives from SYN+ACK or not. Because, the packet is sent from the server to the client via router, hub or any other active device. The server then identifies the reply from the client and stops sending the message to the client and concluded that it is work of a hacker .so the server identifies the SYN flooding attack.

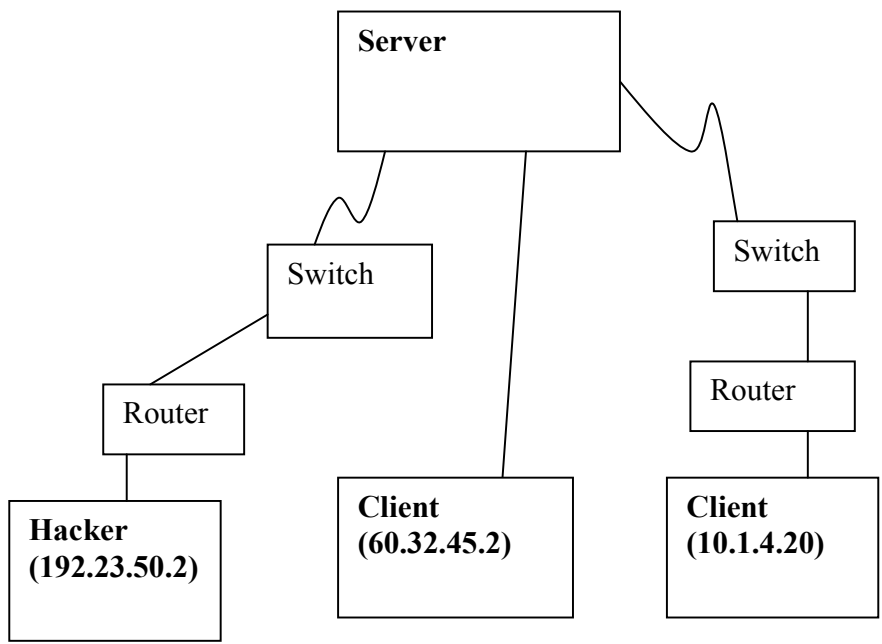

**Fig. 1: Detecting by ICMP Feedback**

# An Active Defense Mechanism for TCP SYN flooding attacks

*Case 4: Detecting by tracing the route*

Another solution is we can trace the route of the corresponding message where it is started from. The ICMP message can be fixed the last router to send original IP address of the router. Now the server can check the source IP address and original IP address whether it is true or not .So the server identifies the SYN flooding attack. In the existing system, there is no time computation mechanism for defense but it is implemented in the proposed system. This defense mechanism is used identify the SYN flooding accurately with short time span and less SYN packets

## 5. CONCLUSION

We have developed an active defense mechanism to protect against TCP SYN Flood attacks using four cases. Our mechanism is based on collaboration among defense nodes through the trace driven approach. SYN attack is detected by combined approach of using TCP flags, port, and ICMP feedback and by tracing the route of the source. Our defense mechanism will be more robust and efficient to detect various SYN flooding attacks. It achieves high detection accuracy and short detection time. Our future work will be to develop ways of identifying attack packets at the hardware level of the system.

## 6. REFERENCE

[1] D. Moore, G. Voelker and S. Savage, August 2001 "Inferring Internet Denial of Service Activity", Proceedings of USENIX Security Symposium'2001.
[2] CERT Advisory CA-1996-21 "TCP SYN flooding and IP spoofing attacks". [Online]. Available: http://www.cert.org/advisories/CA-1996-21.html
[3] CERT advisory CA-1998-01 "smurf IP Denial-of-Service attacks". Available at http://www.cert.org/advisories/CA-1998-01.html, Jan. 1998.
[4] J. Lemon, Feb.2002 "Resisting SYN flooding DoS attacks with a SYN cache," in Proceedings of USENIX BSDCon'2002, pp. 89–98.
[5] A. Zuquete, Sept 2002, "Improving the functionality of SYN cookies," in Proceedings of 6th IFIP Communications and Multimedia Security Conference, pp. 57–77.
[6] RFC793 (rfc793)-Transmission control Protocol. available at www.faqs.org/rfcs/rfc793.html viewed on 1 Sep 2007
[7] Vasilios A. Siris and Fotini Papagalou,2004, "Application of Anomaly Detection Algorithms forDetecting SYN Flooding Attacks", IEEE Communications Society Globecom
[8] BoonPing Lim Md. Safi Uddin, 2005, "Statistical-based SYN-flooding Detection Using Programmable Network Processor". (ICITA'05)
[9] J. Haggerty, T. Berry, Q. Shi and M. Merabti, 2004, "A System for Early Detection of TCP SYN Flood Attacks", IEEE Communications Society Globecom.
[10] H. Wang, D. Zhang, and K. G. Shin, 2002, "Detecting SYN flooding attacks," *IEEE INFOCOM 2002*.
[11] Changhua Sun, Jindou Fan, Bin Liu, Aug 2007, "A Robust Scheme to Detect SYN Flooding Attacks", in *Proc. International Conference on Communications and Networking in China (ChinaCom)*, Shanghai, China.
[12] L. Ricciulli, P. Lincoln, and P. Kakkar, Jan. 1999, "TCP SYN Flooding Defense",. in*Proc. CNDS '99*, pp. 17.20.
[13] T.Peng, C.Leckie,and K.Ramamohanarao, Nov 2002, "Detecting distributed denial of service attacks using source IP address monitoring",
[14] Gemona, I. Duncan, and A. Miller, june 2006,"NEMESI: Using a TCP finite state machine against tcpsyn flooding attacks," in The 6th Annual Postgraduate Symposium on The Convergence of Telecommunications, Networking and Broadcasting (PGNet 2006) pp. 297-303,
[15] Y. Ohsita, S. Ata, and M. Murata, Dec 2004, "Deployable overlay network for defense against distributed SYN flood attacks," Technical Reports of IEICE (IN2003-125), pp. 13–18.
[16] B. Al-Duwairi Duwairi and G. and G. Manimaran Manimaran, 2005, "Intentional dropping: A novel Intentional dropping: A novel scheme for SYN flooding mitigation, scheme for SYN flooding mitigation," in Global Internet in Global Internet Symposium 2005.
[17] Steven M. Bellovin, "ICMP Traceback Messages", Internet Draft: draft-bellovin-itrace-00.txt, submitted Mar. 2000.